\documentclass{article}

\usepackage{arxiv}

\usepackage[utf8]{inputenc} 
\usepackage[T1]{fontenc}    
\usepackage{hyperref}       
\usepackage{url}            
\usepackage{booktabs}       
\usepackage{amsfonts}       
\usepackage{nicefrac}       
\usepackage{microtype}      
\usepackage{graphicx}
\usepackage{doi}

\usepackage{cite}
\usepackage{amsmath,amssymb,amsfonts}
\usepackage{algorithmic}
\usepackage{textcomp}
\usepackage{xcolor}
\usepackage{todonotes}

\title{Uncertainty Annotations for Holistic Test Description of Cyber-physical Energy Systems
}

\author{{Kai Heussen}\thanks{This work has been supported by the ERIGrid 2.0 project of the H2020 Programme under Grant Agreement No. 870620, and publication was supported by the Danish national funding under EUDP IEA 2023-II3, Jrn.No. 134233-511050.}\\
\textit{Department of Wind and Energy Systems} \\
\textit{Technical University of Denmark}\\
Kgs. Lyngby, Denmark \\
0000-0003-3623-1372
\And
{Jan Sören Schwarz, Eike Schulte}\\
\textit{R\&D Division Energy} \\
\textit{OFFIS Institute}\\
Oldenburg, Germany \\
0000-0003-0261-4412, 0000-0002-9078-9791
\And
{Zhiwang Feng}\\
\textit{Institute for Energy and Environment} \\
\textit{University of Strathclyde}\\
Glasgow, United Kingdom \\
0000-0001-5612-0050
\And
{Leonard Enrique Ramos Perez}\\
\textit{European Distributed Energy} \\
\textit{Resources Laboratories (DERlab) e.V.}\\
Kassel, Germany \\
0000-0001-7912-453X
\And
{John Nikoletatos}\\
\textit{Centre for Renewable Energy Sources} \\
\textit{and Savings CRES }\\
Pikermi, Athens, Greece \\
0000-0002-0490-5611
\And
{Filip Pröstl Andrén}\\
\textit{Center for Energy} \\
\textit{AIT Austrian Institute of Technology}\\
Vienna, Austria \\
0009-0007-1809-6740
}

\renewcommand{\shorttitle}{Uncertainty Annotations for Holistic Test Description}

\hypersetup{
pdftitle={Uncertainty Annotations for Holistic Test Description},
pdfsubject={eess},
pdfauthor={Kai Heussen, Jan Sören Schwarz, Eike Schulte, Zhiwang Feng, Leonard Enrique Ramos Perez, John Nikoletatos, Filip Pröstl Andren},
pdfkeywords={design of experiments, geographically distributed real-time simulation, holistic test description, purpose of investigation, uncertainty.},
}

\begin{document}
\maketitle
\begin{abstract}
The complexity of experimental setups in the field of cyber-physical energy systems has motivated the development of the Holistic Test Description (HTD), a well-adopted approach for documenting and communicating test designs. Uncertainty, in its many flavours, is an important factor influencing the communication about experiment plans, execution of, and the reproducibility of experimental results. The work presented here focuses on supporting the structured analysis of experimental uncertainty aspects during planning and documenting complex energy systems tests. This paper introduces uncertainty extensions to the original HTD and an additional uncertainty analysis tool. The templates and tools are openly available and their use is exemplified in two case studies. 
\end{abstract}

\section{Introduction} \label{sec:introduction}

The widespread deployment of distributed renewable energy generators in recent years has fundamentally transformed the planning and operation of electric power systems. To manage these new challenges, automation and control systems leveraging advanced information and communication technologies have become essential. Consequently, smart grid solutions have evolved into complex, multidisciplinary systems. As digitalization progresses and integration with other energy systems expands, new testing scenarios, profiles, and processes must be established.

Traditionally, a significant gap existed between proof-of-concept simulations and prototypical implementations, making the iteration between design and testing slow and costly. However, advancements in real-time simulators (RTSs) have helped bridge this gap \cite{Laus2021}. Additionally, laboratory and field experiments often serve different investigative purposes. While field tests capture real-world operational scenarios, some aspects remain difficult or impossible to analyze comprehensively. Conversely, laboratory experiments enable testing of functionalities that field trials cannot accommodate, particularly where regulatory and security constraints are in place \cite{maniatopoulos_2017}.

With a broad spectrum of testing methods available — from simulations to real-world experimental setups — there is no single "correct" approach, which adds to the complexity. A key requirement for testing smart energy system applications is the reproducibility of test results. Among other factors, various sources of uncertainty make this reproducibility difficult to achieve. The holistic test description (HTD) was developed to offer a standardized framework for defining complex energy system experiments and documenting agreements within a clear context \cite{Heussen2019}, aimed to support stakeholders in decision-making before conducting tests, as well as in documenting the context of completed test results. 
However, in the domain of power and energy systems, no tools are available to systematically address uncertainties during the testing process.

The main contribution of this work is a new tool for structured uncertainty analysis (USAT), and the embedding of uncertainty analysis in the existing HTD and Design of Experiment (DoE) methodologies for test planning.

This paper is structured into five sections. Section \ref{sec:related_work} and \ref{sec:concepts} introduce the background work and relevant concepts related to HTD, design of experiments (DoE) as well as the uncertainty concept applied to test design and planning. Section \ref{HTD-UA} elaborates on the aspects involved in the uncertainty quantification (UQ) as part of the HTD. Then, Section \ref{sec:application_cases} presents two test cases on cyber-physical energy systems in which both the HTD improvements and the experimental uncertainty analysis tool are implemented. Section \ref{sec:discussion} concludes with an outlook toward further automation of the presented approach.

\subsection{Related Work} \label{sec:related_work}

The HTD methodology is described in \cite{Heussen2019}. A first step toward methodical application of the Design of Experiments (DoE) to HTD was achieved by developing a structured guideline, reported in~\cite{8405401}. 

A dedicated repository at GitHub\footnote{\url{https://github.com/ERIGrid2/Holistic-Test-Description}} stores the templates and guidelines related to HTD. It contains guidelines for describing test cases (from a holistic approach), identifying test components, purpose of investigation and qualification strategy. Several examples of the HTD use are found here. A repository with HTD example cases is found here, and a method for profiling test cases is proposed in \cite{Raussi2023_tcprofiles}. A python toolbox for DoE for simulation and hardware experiments is proposed and showcased in \cite{schwarz_2024_toolbox}, which is suitable for implementing uncertainty analyses identified by the methods presented here.




\begin{figure}[t]
	\centering
	\includegraphics[width=1\linewidth]{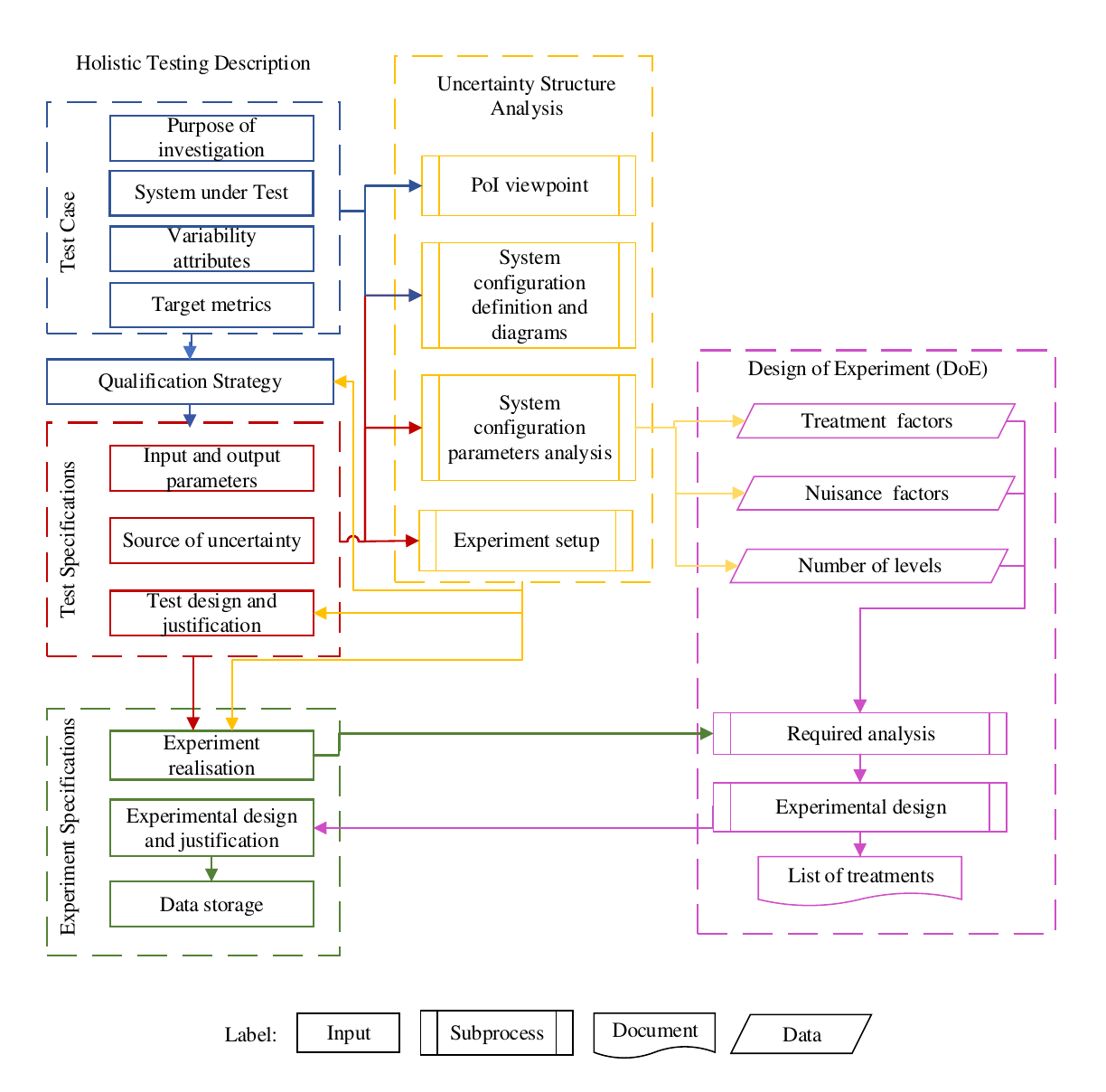}
	\caption{DoE as a part of the holistic testing workflow. From \cite{8405401}   }
	\label{fig:DoE_HTD}
 \end{figure}

\section{Background Concepts} \label{sec:concepts}

\subsection{Holistic Test Description - Key Concepts} 
The main goal of the HTD methodology is to provide tools that support users in creating and documenting tests of smart grid components and systems. During the process relevant questions are raised helping users in the early phases of test scoping and planning ~\cite{Heussen2019}.

For this purpose, the HTD methodology is based on three main stages, which are also visualised on the left-hand side of Figure \ref{fig:DoE_HTD}. First, a test case is created providing a general description of the system under test with its main functions. The test case also includes the objective of the investigation and the test criteria. The second stage of the HTD is used to define specific configurations, inputs, outputs, metrics and test protocol. The third stage, called experiment specification, covers both the physical experiment set up and lab process. HTD offers templates to elicit relevant information throughout the full process, which thus is directly documented. The templates that enable a structured and simplified communication between the stakeholders involved in the testing process.

In the original version of the HTD, uncertainty aspects where only partly covered. Although it helped the users to identify sources of uncertainty, and in mapping those to an analytical Design of Experiments (DoE) approach, as illustrated in Fig.~\ref{fig:DoE_HTD} (violet boxes). Here e.g. HTD "variability attributes" could be mapped to DoE "nuisance factors"; however, no appropriate tools were provided to manage the wide variety of uncertainties that emerge in testing \cite{8405401}.


\subsection{About Uncertainty in Test Preparation.}
Three types of sources of uncertainty in preparing complex test setups can be distinguished: \textit{(i.)} Uncertainty about realization of intended test on (foreign) laboratory infrastructure; \textit{(ii.)} Uncertainty about the problem statement; and \textit{(iii.)} the behavioural and quantifiable uncertainty for a given experimental hypothesis. Of the above uncertainty Aspects, \textit{(i.)} and \textit{(ii.)} can be addressed by working with HTD already, which strongly facilitates the multi-perspective analysis of a testing problem.
However, the mathematical complexity of correctly addressing uncertainty in an experimental design, aspect \textit{(iii.)}, can be overwhelming, since any representation is \textit{per-se} uncertain, there are usually uncontrollable influences in the experiment, both intended and unintended ones, such as from the test equipment and instrumentation. 
\subsection{Uncertainty Analysis, Representation and Quantification}

Underlying the proposed approach for integration of uncertainty into HTD is an analysis considering \textit{uncertainty framing}, \textit{mathematical representation}, and \textit{quantification methods}; which has been further detailed in \cite{jra_1_2}.
\subsubsection{Uncertainty Framing}
Uncertainty can be classified as aleatory or epistemic \cite{bib:KIUREGHIAN2009}.
The correct classification is an important first step to find a suitable approach to handle it.
Aleatory uncertainty results from intrinsic fluctuations of a system and is not reducible.
Examples include the irradiation of the sun and random measurement errors in the context of hardware-based experiments.
Aleatory uncertainty leads to results that are not exactly reproducible, meaning that experiments have to be repeated multiple times and analysed with statistical methods.
%
Epistemic uncertainty on the contrary, results from lack of knowledge. 
Thus, one option for reducing it can be gaining more knowledge.
If that is not possible, statistical analysis might help as well.
For example, for a deterministic software-based experiment no intrinsic randomness occurs, but the parameterization might not be exactly known.
In this case, a sampling-based approach can help to investigate how changes in the parameters affect the results.
\par

\subsubsection{Uncertainty Representation}
To manage uncertainty, one has to find a good representation for describing it.
For representation of aleatory uncertainty, probability theory is usually used and uncertainty is described in the form of probability distributions.
In some cases, distributions can also be used for epistemic uncertainty, but other types of representations might be better suitable, taking into account that the distribution parameters are not known. 
Commonly, epistemic uncertainty is addressed by defining expected intervals for uncertain parameters, but more complex representations like 
probability boxes, Dempster-Shafer theory, possibility theory, or fuzzy sets can also be used \cite{jra_1_2}.
\subsubsection{Uncertainty Quantification}
The main task of \textit{uncertainty quantification (UQ)} is to measure the impact of uncertainties on the input variables (factors) on the output of the system, which can be handled with different approaches (see also Table 3 in \cite{jra_1_2}).
The \textit{analytical approach} propagates the uncertainty based on a simple mathematical description of the system, best used for smaller systems or rough approximations.
In contrast, \textit{sampling approaches} consist of repeating an experiment multiple times with changing input values, often used for UQ or \textit{sensitivity analysis (SA)} \cite{Zhang2020}. Accurate UQ can also be performed by explicitly propagating the uncertainty through a system, as proposed for stationary simulation models in \cite{bib:Steinbrink2017}.
In support of reducing the number of factors to assess by computationally expensive UQ, SA can be used to rank importance of factors for prioritizing those with largest effects on the outcome. SA can be differentiated into \textit{local SA} which focusses on exact effect ranking over a small region of the parameter space, and \textit{global SA}, which enables a ranking over the full range of the parameter space, as illustrated in \cite{schwarz_2024_toolbox}.
%



\section{HTD with Uncertainty Analysis} \label{HTD-UA}




\subsection{HTD Uncertainty Structure Analysis Tool (USAT)}\label{HTDUSATool}

\begin{figure*}[t]
	\centering
	\includegraphics[width=1\linewidth]{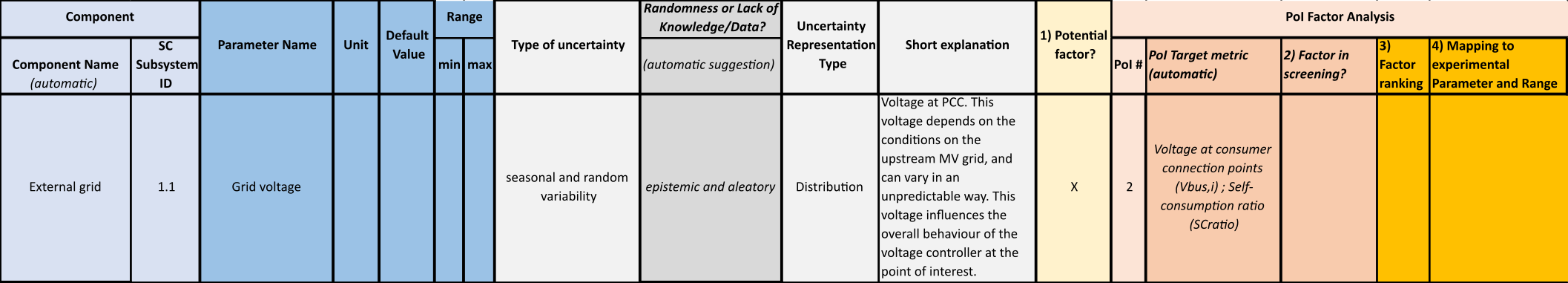}
	\caption{‘SC Parameter Analysis’ sheet of the HTD USAT. }
	\label{fig:USAT_SC_Parameter}
 \end{figure*}

To support the user in annotating and handling the previously mentioned types of uncertainties in the HTD, a new tool was developed and the HTD template extended as described as follows.
The HTD USAT is provided in form of a Microsoft Excel template\footnote{\url{https://github.com/ERIGrid2/holistic-test-description/tree/master/UncertaintyStructureAnalysis}} 
, facilitating structured analysis and documentation of the uncertainty aspects. 
The USAT Tool addresses the interrelated uncertainty aspects across all HTD layers, as illustrated in the yellow boxes and arrows in Fig. \ref{fig:DoE_HTD}:
\begin{itemize}
    \item Uncertainty assessment methodology aspects for each \textbf{Purpose of Investigation (PoI)}, with space for definitions of target metrics automatic references to identified factors;
    \item An object-oriented breakdown of the \textbf{System Configuration (SC)} defining its components and subsystems, using a system-breakdown diagram (SBD) approach;
    \item A worksheet based on listing \textbf{SC parameters}, cf. blue area in Fig.~\ref{fig:USAT_SC_Parameter}, where type of uncertainty, framing and representation type can be identified for each parameter (grey area); the yellow/orange columns support factor selection and ranking approaches.
    \item Finally, an analysis tool for uncertainty aspects in the \textbf{Experiment Setup (ES)} includes different types of ES (e.g., software-based, hardware-based, mixed), which can be chosen and typical uncertainty aspects are proposed for further consideration.
\end{itemize}



%


\subsection{Using the HTD USAT} \label{sec:using_USAT}

As a preparatory phase, to prepare the data before the actual execution of the test procedure, the definition of the goals of the experiment and the identification of the outputs to be measured or calculated are crucial for the process of uncertainty quantification (UQ) and are part of the HTD.
Thus, the defined PoI and target metrics are transferred from the HTD template to the ’PoI viewpoint’ sheet of the USAT. Depending on the defined PoI, specific points of the identification process are emphasized. The viewpoint differentiates depending on the specific objective of interest, in terms of uncertainty, sensitivity, or scaling analysis.
\par
Further, the factor analysis for the identification of the factors that may potentially influence the outputs, being thus sources of uncertainty, is based on the information of the ‘SC definition and Diagrams’ sheet of the USAT.
Here, the SBD representation of the System Configuration (SC) can be utilized. The SBD diagram is transferred in the tool and its components are described. The SBD provides an organised overview of all components that constitute the SC and facilitates the procedure for the systematic identification of sources of uncertainty. The sources of uncertainty are explored by examining the system components, one by one, and identifying the parameters that may potentially disturb the outputs, quantities of interest, of the system. This task is based on the experience and knowledge of the practitioner. The identified parameters are characterised with the aid of a set of predefined taxonomy terms. In addition, a type for the mathematical representation of the uncertainty may be suggested, as well as a range of the variation of the values of the parameters. 

The final outcome of this process, aims to cover the identification of the variable attributes of the TC level of the HTD. Each factor identified by the aforementioned procedure is assigned to a PoI case, and the list of the selected factors for each assigned PoI case is available and displayed in the ‘PoI viewpoint’ sheet. 

For complex test systems, typically a large number of factors can be identified, yet only a few of these have a significant effect on the model output. Screening experiments help to identify the subset of factors that control most of the output variability with a relatively low computational effort. 
The information on the ranking of the examined factors can then be integrated into the HTD USAT. The tool thus collects in a single document all the main uncertainty aspects from the preliminary analysis, and can facilitate subsequent methods of UQ, which may require more detailed analysis. 
%
%
%
%
%

\subsection{Uncertainty Extensions for HTD Template}
The basic HTD template 
 was amended to persistently consider uncertainty aspects of technology, system configuration and lab properties in the following aspects (see \cite{na_4_3} for more details):

\begin{itemize}
    \item Test Case (TC): Adds a new field for detailed PoI and factor analysis alongside existing variability and quality attributes.
    \item Qualification Strategy (QS): Now includes uncertainty identification and management strategy.
    \item Test Specification (TS): Merges input/output parameters with sources of uncertainty, linking to the detailed uncertainty structure analysis
template.
    \item Experiment Realisation: Links to USAT “ES viewpoint” sheet for assessing uncertainty trade-offs.
    \item Experimental Specification (ES): Adds new fields for experimental setup uncertainties, precision of equipment, measurement uncertainty, and uncertainty management, and is also linking to USAT “ES viewpoint”.
\end{itemize}

\begin{figure*}[t!]
	\centering
	 \includegraphics[width=0.8\linewidth, trim={0.0cm 6.65cm 0.0cm 0cm}, clip]{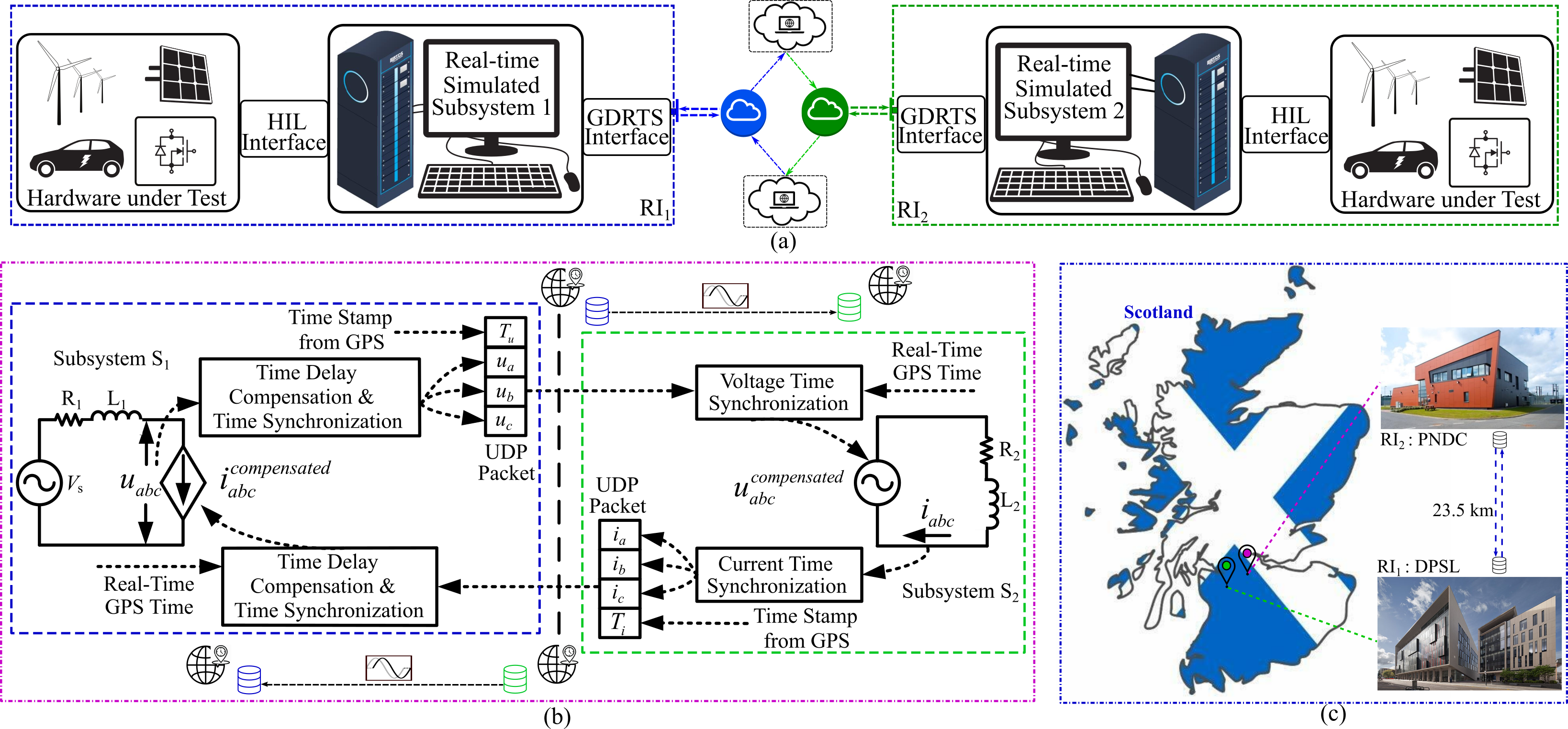}\\
     \footnotesize{(a)}
	 \includegraphics[width=1\linewidth]{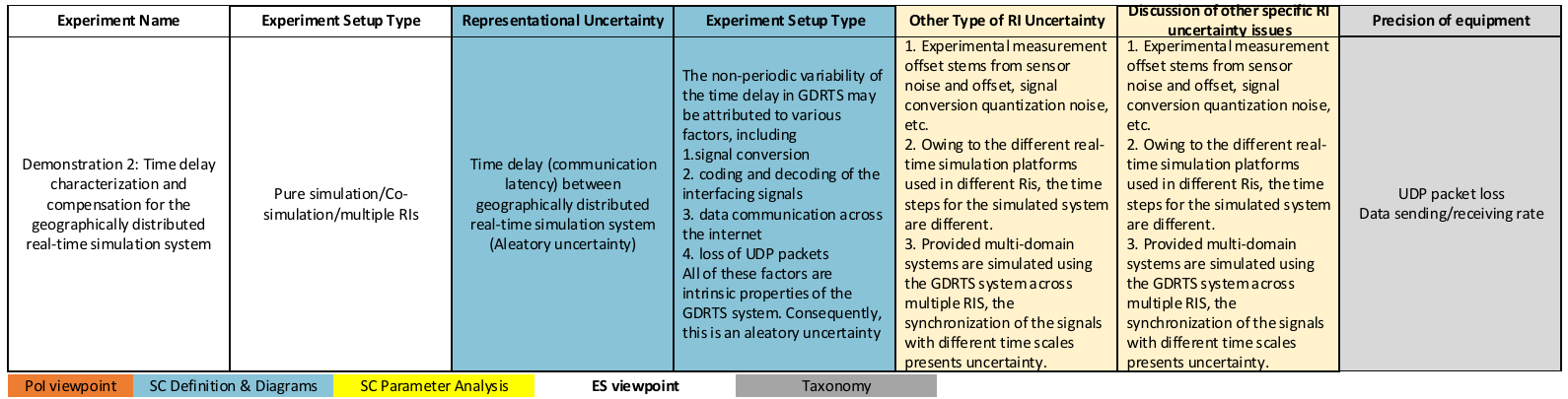}
     \footnotesize{(b)\\}

	\caption{(a) Representation of GDRTS-PHIL test setup with two RIs and exemplary hardware under test equipment and where communication time delay will affect the GDRTS interface, (b) the associated analysis in experiment specification (ES) viewpoint of the USAT.}
	\label{fig:GDRTS-diagram}
 \end{figure*}

\section{Application Cases} \label{sec:application_cases}

\subsection{Uncertainty Screening of a Multi-Energy Co-Simulation}
%
The utilization of HTD and USAT for facilitating the identification and classification of uncertainties is illustrated in an example of a test case related to the multi-energy network benchmark (MENB) developed in ERIGrid~2.0\cite{ait_ces_2021_5735005,DJRA101}. 
\subsubsection{\textbf{Test Case Description}}
This benchmark model contains a coupled heat and power network, and the specific test case addresses issues related to self-consumption of excess power generated from renewable (PV) power generators. 
An electrical and a thermal network supply power and heat to their connected consumers. The networks are coupled through a power-to-heat facility comprising a heat pump connected to a thermal tank, which feeds into the thermal network’s supply line to support its operation.
The electrical energy generated by the PV systems connected to the electrical network and the operation of the heat pump for the conversion to thermal energy and storage to a hot water tank can be coordinated for a more efficient use of the networks. In particular, the operation of the heat pump is adjusted according to the imbalance between PV generation and consumption, as determined by the voltage deviations in the local electrical network. The local use of excess power helps mitigate voltage variation problems in the electric distribution network, improving at the same time the self-consumption.

The system  operation is determined by two controllers: 

 A \textbf{voltage controller} monitors the voltage in the electrical network and proposes a power consumption set point for the heat pump to keep the voltage within acceptable limits. 

The \textbf{heat controller} operates the power-to-heat facility and decides whether the power-to-heat facility supports the thermal network by discharging the tank. 

This dependency of the sub-systems causes a complex interaction which can only be fully captured by assessing them simultaneously.

\subsubsection{\textbf{Uncertainty Parameter Screening Using the HTD USAT}}

Based on the MENB test case the procedure for the utilization of the USAT within the HTD process is presented, with visual samples of the tool after having been filled in a practical example. A filled HTD template of the multi-energy network benchmark was already available and was used as a basis to fill out the USAT\footnote{\url{https://github.com/ERIGrid2/holistic-test-description/tree/master/Examples/Multi-Energy-Benchmark}}.
Following the process described in Section \ref{sec:using_USAT}, the defined PoI and target metrics have been transferred to the’ PoI viewpoint’ sheet of the Excel tool, differentiated depending on the specific objective of interest. In this example, the System Breakdown (SBD) representation of the System Configuration (SC) is utilized. 
The SBD diagram has been transferred in the tool and its components have been described. Then the sources of uncertainty are systematically identified and characterised, by examining the system components, one by one. A view of the ‘SC Parameter Analysis’ sheet, after this procedure, is shown in Figure \ref{fig:USAT_SC_Parameter}, with a part of the list of the identified factors. The characterisation and the assignment to each PoI are included also for each factor, besides other details.

In the HTD USAT, the factors for SA were marked and a first rough SA with an One-at-a-Time (OAT) approach was adopted for a preliminary ranking of the factors. For the implementation a toolbox for SA and DoE was used, which is described in detail in \cite{Schwarz2023}. The information on the ranking of the examined factors has been integrated in the HTD USAT. Based on the resulting ranking of the impact factors have on the target metrics, some factors were chosen which might be considered in a more detailed analysis.  


\subsection{Uncertainty Analysis for Delay Characterization Scenario} 


In this case study, the HTD template with uncertainty extensions and the HTD USAT depicted in Section \ref{HTD-UA} were employed to facilitate the uncertainty identification and analysis in the test case involving the time delay characterization of the geographically distributed real-time simulation (GDRTS) setup in the ERIGrid 2.0 project.

\subsubsection{\textbf{Test Case Description}} Real-time coupling and integration of geographically distributed RI via the GDRTS approach has been developed to leverage the equipment, real-time simulations, and expertise of individual RI collaboratively to address the evolving needs for complex energy system experimental validation. As illustrated in Fig.\,\ref{fig:GDRTS-diagram}(a), the data and signals within the GDRTS setup are transmitted over the internet through the dedicated interface that bridges multiple RIs into a closed-loop configuration. The indeterministic nature of power signal data exchange over the internet and signal conversion units introduces substantial uncertainty about the time delay in the GDRTS closed-loop setup as depicted in \cite{jra_1_2}. From an application perspective, the time delay with time-varying yet deterministic attributes can be regarded as an uncertainty factor that affects the fidelity of GDRTS operation. Having a comprehensive uncertainty representation and quantification of the time delay via the proposed HTD template and the HTD USAT is of great significance for accelerating the design and deployment of the GDRTS setup. This will facilitate a more comprehensive evaluation of geographically distributed complex energy systems and power apparatus with uncertainty factors taken into account.

\subsubsection{\textbf{Uncertainty aspects for HTD and USAT aided uncertainty analysis}} According to the specifications detailed in Section \ref{HTD-UA} regarding the HTD with uncertainty extensions and technical integration, the uncertainty representation of the GDRTS time delay characterization test case was accomplished by utilizing the proposed HTD template\footnote{\url{https://github.com/ERIGrid2/holistic-test-description/tree/master/Examples/GDRTS_Time-Delay_Case}}. The HTD template consists of the following:
\begin{itemize}

\item \textbf{\textit{PoI viewpoint} sheet}: The PoI (i.e., the GDRTS time delay characterization) of this test case along with the explanation of where the uncertainty arises from, are depicted in this sheet. The target metrics of the system signal phase error and power error stemming from the time delay are presented.  Based on the subsequent sheets in this HTD template, the factors in screening and their ranking are also presented.

\item \textbf{\textit{SC Definition\,\&\,Diagrams} sheet}: Initially, the GDRTS system depicted in Fig.\,\ref{fig:GDRTS-diagram}(a) was represented in the form of SC by using the SBD criteria. As shown in this sheet, the SC was represented by dividing the holistic GDRTS system into multiple subsystems, each with a dedicated name, ID number, along with a short description.

\item \textbf{\textit{SC Parameter Analysis} sheet}: Based on the well-defined SC, the details of the components leading to system uncertainty in each SC subsystem are further depicted in this sheet. Upon developing this sheet, the source of uncertainty of these components and their types (aleatory or epistemic) are further categorized. Furthermore, the uncertainty representation type for each component is outlined for subsequent uncertainty quantification.  

\item \textbf{\textit{ES viewpoint} sheet}: As illustrated in Fig.\,\ref{fig:GDRTS-diagram}(b), the different types of uncertainties for this GDRTS time delay characterization test case are presented from the ES perspective. Note that, the communication latency has been categorized as representational uncertainty. 
\end{itemize} 

Based on the above sheets in the HTD template, the HTD USAT is leveraged to screen the uncertainty factors listed in the \textit{SC Parameter Analysis} sheet to identify and rank the subset factors according to their impact on the GDRTS time delay. Subsequently, the information on the ranking of the examined factors is listed in the \textit{PoI viewpoint} sheet. By doing so, the tool thus consolidates all the main uncertainty aspects with their respective details into a single document. It is evident that communication latency is ranked as the predominant factor impacting the GDRTS time delay. This information is critical for facilitating the subsequent UQ and UA, thereby supporting the planning and documentation of the GDRTS system. 

 \begin{figure}[t!]
	\centering
	\includegraphics[width=0.95\linewidth,trim=0.6cm 0.4cm 0.8cm 0.0cm,]{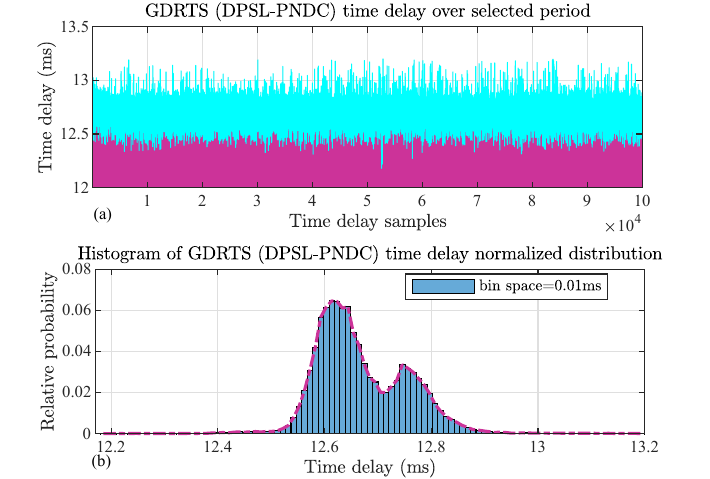}
    \vspace{.5cm}
 
	\caption{Representation of the GDRTS time delay between DPSL and PNDC.}
	\label{fig:GDRTS-delay_charac}
 \end{figure}
 
\subsubsection{\textbf{Uncertainty Representation \& Quantification}} As highlighted in the output of the HTD template, the communication latency emerges as the most critical uncertainty factor to be quantified. Consequently, experimental delay data across multiple RIs were collected between the Dynamic Power Systems Laboratory (DPSL) and the Power Network Distributed Center (PNDC), which are located approximately 23.5\,km away from each other geographically at the University of Strathclyde in Scotland. To represent the uncertainty of the delay within this GDRTS setup, the aggregated loop delay data between the geographically distributed DPSL and PNDC labs with 100\,k delay samples over a randomly selected period was presented in Fig.\,\ref{fig:GDRTS-delay_charac}(a). It is evident that the loop delay varies over time with non-periodic characteristics. The non-periodic variability of the time delay in this GDRTS may be attributed to various factors as listed in the HTD template. The non-periodic and time-varying delay is a critical uncertainty factor that necessitates thorough quantification. 

In an effort to quantify the uncertainty of the delay within the GDRTS setup, the statistical analysis methods were employed to process the delay data presented in Fig.\,\ref{fig:GDRTS-delay_charac}(a). At the initial stage, the delay data was grouped into 100 bins that are evenly distributed across the range of delays collected within the GDRTS setup. The range of delays spans from 12.18\,ms to 13.20\,ms. The relative probability is denoted as $\rho_i$ (i.e., ${c_i}/{N}$) for each delay bin $i$, with $c_i$\,-\,the cumulative number of observations for each delay bin, and $N$\,-\,the total number of delay samples. Furthermore, $\rho _i$ is normalized by dividing the cumulative number of observations for each delay bin by $N$. This provides insights into the underlying numerical distribution of delays and their frequency of occurrence and allows us to illustrate the relative probability of a delay value falling within each delay bin, as in Fig.\,\ref{fig:GDRTS-delay_charac}(b). 

The relative probability of the variable delay falling within the bin edges $[12.60\,ms, 12.61\,ms]$ is the highest, reaching a value of 6.46\,\%. On the other hand, the relative probability of the variable delay falling within the bin edges at the lowest and greatest values of the range, i.e., $[12.18\,ms, 12.19\,ms]$ and $[13.19\,ms, 13.20\,ms]$, respectively, is minimal, measuring at 0.001\,\% and 0.003\,\%, respectively. This reveals the time-varying and nonuniform characteristics of GDRTS delay in a probabilistic manner, which aids the assessment of the impact of the delay variability and its associated uncertainty on the GDRTS accuracy. Moreover, this UQ with nonuniform probability distribution modeling will be further applied to the UA as outlined in Section 5.3 in \cite{jra_1_2}.

\section{Discussion and Conclusion} \label{sec:discussion}

Uncertainty handling in experiment design is a critical component in enabling reproducibility. The presented HTD extensions and the USAT tool pave the way for clearer documentation, but also for well-structured softwere tooling in order to manage complex CPES experiments. USAT could be further automated and integrated with actual experiment execution. For example, it could be integrated with a software toolbox for DoE integration \cite{schwarz_2024_toolbox}, which also uses an object-oriented SC model and automatically performs sample generation and integrates the appropriate statistical analysis. 


\bibliographystyle{IEEEtran}
\bibliography{literature}

\end{document}